\documentclass[a4paper,twocolumn,superscriptaddress,pre, aps, floatfix ]{revtex4-1}
\usepackage{amssymb}
\usepackage{amsmath}
\usepackage{marvosym}
\usepackage{graphicx}
\usepackage{color}
\usepackage{url}
\usepackage{lipsum}
\usepackage{soul}
\usepackage{multirow}
\usepackage[colorlinks=true,allcolors=blue]{hyperref}
\usepackage{natbib}

\begin{document}
%
%
%
% =================================================================================================
% HEADER
% =================================================================================================
%
\title{
The Physics of Fun: Quantifying Human Engagement into Playful Activities}
\author{David \surname{Reguera}}

\affiliation{Departament de F\'isica de la Mat\`eria Condensada, Universitat de Barcelona, Mart\'i i Franqu\`es 1, E-08028 Barcelona, Spain}
\affiliation{Universitat de Barcelona Institute of Complex Systems (UBICS), Universitat de Barcelona, 08028 Barcelona, Spain}

\author{Pol \surname{Colomer-de-Sim\'on}}

\affiliation{Departament de F\'isica de la Mat\`eria Condensada, Universitat de Barcelona, Mart\'i i Franqu\`es 1, E-08028 Barcelona, Spain}

\author{Iv\'an \surname{Encinas}}

\affiliation{King Digital Entertainment, 08029 Barcelona, Spain}

\author{Manel \surname{Sort}}

\affiliation{King Digital Entertainment, 08029 Barcelona, Spain}

\author{Jan \surname{Wedekind}}

\affiliation{King Digital Entertainment, 08029 Barcelona, Spain}

\author{Mari\'an \surname{Bogu\~n\'a}}

\affiliation{Departament de F\'isica de la Mat\`eria Condensada, Universitat de Barcelona, Mart\'i i Franqu\`es 1, E-08028 Barcelona, Spain}
\affiliation{Universitat de Barcelona Institute of Complex Systems (UBICS), Universitat de Barcelona, 08028 Barcelona, Spain}

\date{\today}
%
%\pacs{}
%
\begin{abstract}
  Engaging in playful activities, such as playing a musical instrument, learning a language, or performing sports, is a fundamental aspect of human life. We present a quantitative empirical analysis of the engagement dynamics into playful activities. We do so by analyzing the behavior of millions of players of casual video games and discover a scaling law governing the engagement dynamics. This power-law behavior is indicative of a multiplicative (i.e., “happy- get-happier”) mechanism of engagement characterized by a set of critical exponents. We also find, depending on the critical exponents, that there is a phase transition between the standard case where all individuals eventually quit the activity and another phase where a finite fraction of individuals never abandon the activity. The behavior that we have uncovered in this work might not be restricted only to human interaction with videogames. Instead, we believe it reflects a more general and profound behavior of how humans become engaged in challenging activities with intrinsic rewards.
\end{abstract}
\maketitle
%
%
%
%
%
% =================================================================================================
\section{Introduction}
% =================================================================================================
%

Humans are deeply captivated to try new experiences that eventually become pleasant daily routines. The enjoyment of playing a musical instrument, speaking foreign languages, sports or hobbies, are all activities that for full enjoyment require some time investment and training experience that eventually pay off. One interesting question is how humans get engaged and come to love these activities, which offer both a challenge as well as an intrinsic reward. What is the training or learning process and how does it affect their level of enjoyment? How can we measure and quantify fun? 

Before the new era of modern technology, answering this type of question relied on the accumulated knowledge obtained from qualitative observations of single individuals made in different conditions by different observers. This makes it very difficult to extract general laws of human behavior. The widespread use of the Internet and the world-wide connectivity that it provides is changing this picture radically and fast. For the first time in human history, it is possible to monitor human actions on an unprecedented large-scale, allowing us to uncover precise and quantitative laws of human behavior~\cite{Lazer721,Conte2012,CHANG201467}. Nowadays, we have the ability to measure, with impressive precision, our mobility patterns~\cite{Gonzalez:2008aa,Song1018}, our musical tastes~\cite{Buldu:2007aa,Celma2010}, or the way in which ideas spread and crystallize across populations~\cite{Centola1194,Bakshy:2011aa}, providing us with a very accurate picture of some of the key aspects of human behavior at the large scale~\cite{Barabasi2005,Rybski2009,Brockmann2006,Liljeros2001}.

Fostered by the widespread outburst of smart phones and tablets, one of the most popular current amusements are casual video games. These are games with simple rules and game dynamics that can be played in brief bursts in a casual way, e.g. during breaks or daily commuting. Some of these games, like Candy Crush Saga (the flagship game of King Digital Entertainment), have reached outstanding popularity. As of the fourth quarter of 2018, King’s games were played by 268 millions monthly active players, with millions of players playing many millions of levels every day in Candy Crush Saga alone~\footnote{Activision Blizzard Announces Fourth-Quarter 2018 and Financial Results. Available at: https://investor.activision.com/news-releases/news-release-details/activision-blizzard-announces-fourth-quarter-and-2018-financial.}. Hence, they are an ideal platform for studying how humans become engaged in a rewarding activity.

There is a vast literature on measuring video game engagement and enjoyment\cite{Boyle2012}. However, most of these studies are based on 1) surveys with a moderate number of individuals~\cite{OBrien:2008aa,OBrien:2010aa,Hoffman2010,Olson:2010aa, Charlton2007, Wiebe2014},  2) physical measures of behavioral~\cite{Sharek2014,Yannakakis2007} or physiological metrics (e.g. heart rate, blood pressure, galvanic skin response or electroencephalogram signals) on players while they are playing~\cite{Chanel2011,Liu2009,Yannakakis2008}, or 3) studies of psychological motivations~\cite{Yee:2006aa,Przybylski:2010aa,Olson:2010aa,Ferguson2013}. In this paper, it is not our intention to enter into the psychological, motivational, behavioral, or social aspects of video game playing nor criticize the standard psychometric, behavioral or physiological metrics, or questionnaire-based evaluation of engagement performed on a limited number of individuals (typically aware to be subject of study) and short time span. Our work is radically different as it approaches the problem from a data-driven point of view by analyzing the real behavior of a large population of individuals as they play the game. In some of the games we have analyzed, we follow the individual behavior of a cohort of 10 million players during a period of two years. This astonishing amount of data allows us to quantify empirically users’ engagement vs progression in a way that has not been possible before the big data era. 

Specifically, we show that the progression, engagement, and quitting of players in casual games can be analyzed and simulated using a simple stochastic model. The level of engagement of a fun activity, like a video game, can be measured and shows a common scaling behavior described by a power-law as a function of the progression in the game. This result suggests that enjoyment, like popularity, wealth, and many other phenomena, is a multiplicative process~\cite{Yule1925,Simon1955,Price1976,Barab&aacute;si1999}: the more you are into it, the more engaged you become. Besides, our analysis reveals that this power-law behaviour is universal across many different games, player segmentation, or countries.
Our empirical findings have interesting implications not only for casual games but also for generic engagement dynamics into a variety of different activities, reflecting a global trend of human behavior.
\begin{figure}[t] 
\includegraphics[width =\linewidth]{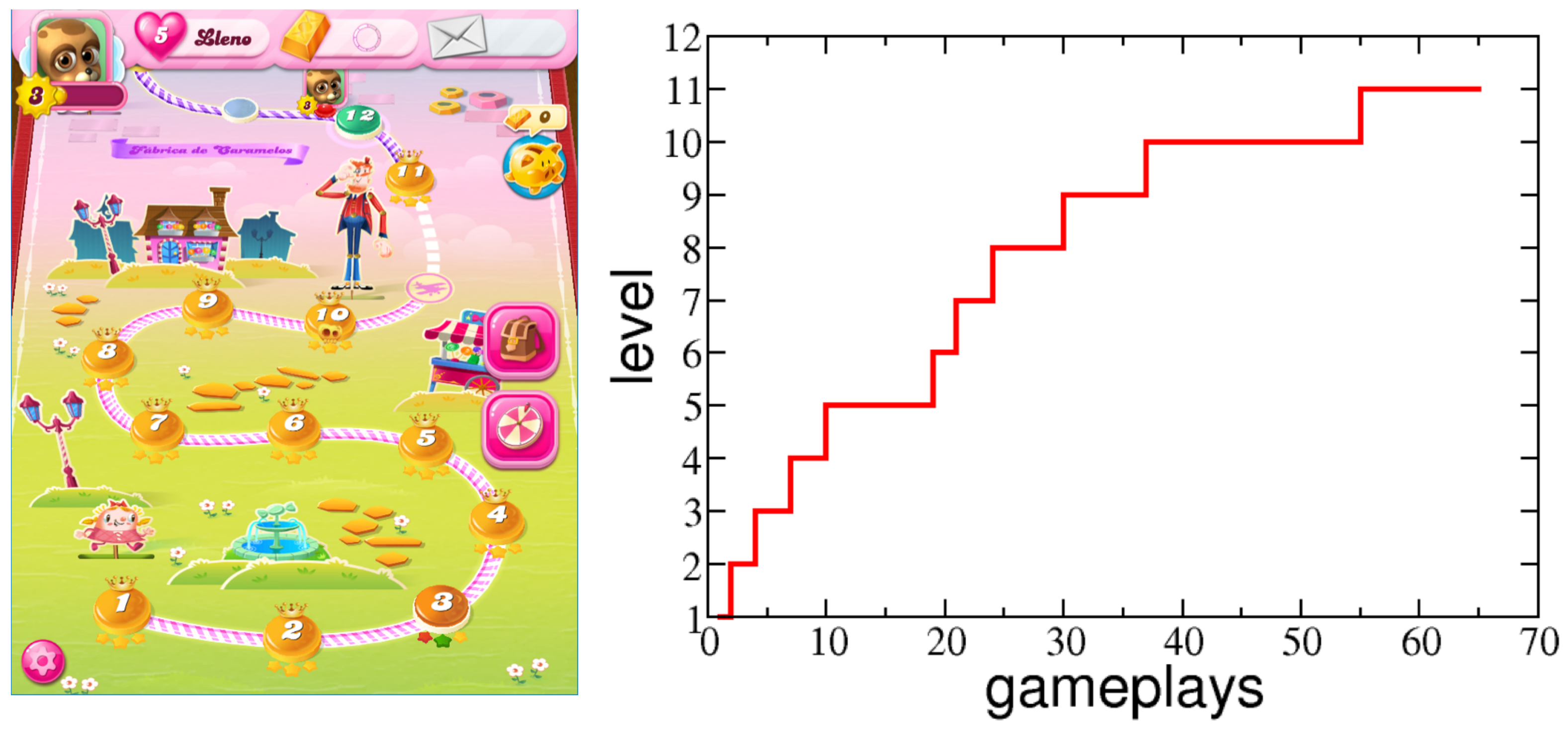}
\caption{(left) Map of the linear sequence of levels of Candy Crush Saga game. Players start at level 1 and take a different number of attempts to pass each level, progressing until eventually they decide to abandon the game. (right) Typical trace of the progression of a player measured as the highest level achieved after a total number of accumulated attempts.}
\label{fig:1}
\end{figure}

\section*{Results}
{\bf Casual Games.} Many typical casual games, like Candy Crush Saga, pose a linear sequence of levels that a player can access one by one as the previous level is successfully completed (see Fig.~\ref{fig:1}). Players start the game at level 1 and progress level by level in an increasing manner. At each level, the player must achieve a predefined goal to pass it (e.g. collect a specific number of candies or reach a certain score) using a limited number of moves, resources, or time. Each attempt to pass a level is a called a “gameplay”. It can be successful, meaning that the player passes that level and can play the next one, or unsuccessful. Alternatively, the player can become tired or frustrated at some point and decide to quit the game. Each level always involves randomness, either in the initial configuration or in the dynamics. This makes it natural to model game dynamics as a stochastic process~\cite{Gardinerbook,vankampen}.  

To model player progression and experience in the game, we use two general indicators: one to quantify the total time spent in the game and another one to measure the progression within the game. In casual games, the real-time activity (i.e. how often and for how long the person plays the game) is not a good measure of the actual time spent in the game. This is so because these games are very often played in short breaks or free time, which is unpredictable and not controlled by the player. Instead, we use the accumulated number of attempts (or gameplays) as activity-independent measure of the total “time” spent in the game.  The maximum level achieved after a given number of gameplays is an indicator of game progression, i.e., on how far a player is in the game (see Fig.~\ref{fig:1}b). With this strategy, we monitor the actual progression of players in the game decoupled from their real-world activity. 

For these games, the dynamics of game progression can be modeled in a very simple way using Continuous Time Random Walks (CTRW)~\cite{Montroll:1965aa,Weiss2007}, as described in detail in the Supplementary Information (SI). In our model, we assume that all players can be considered as identical and independent.  When a player reaches a new level, there are two competing random processes taking place simultaneously~\footnote{We do not consider paying players (i. e., players that pay to pass levels) in this analysis.}: 1) the random number of attempts required to pass that level, $\tau_p$, and 2) the random time, measured in number of attempts, that the player takes to get bored or frustrated and decides to abandon the game, $\tau_a$. For a given level, the final fate of the player depends on which of these random times is shorter. If $\tau_p<\tau_a$, the player passes the level and jumps to the next one; otherwise the player quits the game. These two times are assumed to be statistically independent random variables with probability density functions $\psi_n^p(t)$ and $\psi_n^a(t)$. In short, $\psi_n^p(t)$ controls the time that the player would take to pass level $n$ if he/she were not allowed to abandon the game. Similarly, $\psi_n^a(t)$ defines the time the player would take to abandon the game if level $n$ were impossible to pass (without the player knowing it). In the simplest version of the model, pass and abandon times at level  are taken to be Poisson point processes and, therefore, their probability density functions are~\cite{renewal}
\begin{equation}
\psi_n^p(t)=\frac{1}{\bar{t}_p(n)}e^{-t/\bar{t}_p(n)}, \; \; \psi_n^a(t)=\frac{1}{\bar{t}_a(n)}e^{-t/\bar{t}_a(n)},
\label{eq:1}
\end{equation}
where $\bar{t}_p(n)$ and $\bar{t}_a(n)$ are the average time to pass or abandon at level $n$, respectively~\footnote{In the SI file, we show that although this assumption is not totally correct, deviations from the exponential hypothesis only occur in around $0.1\%$ of the observed abandon and pass times.}. With this choice, $\bar{t}_p(n)$ and $\bar{t}_a(n)$ are the main ingredients of the model. Specifically, $\bar{t}_p(n)$ is a measure of the relative difficulty (or relative cost) of that particular level, whereas $\bar{t}_a(n)$ is a measure of the engagement of a player at that particular level. Both times can be easily measured for an arbitrary dataset as (see Methods)
\begin{equation}
\bar{t}_p(n)=\frac{\bar{t}_p^{emp}(n)}{1-p_c(n)} \mbox{ and } \bar{t}_a(n)=\frac{\bar{t}_p^{emp}(n)}{p_c(n)},
\end{equation}
where $\bar{t}_p^{emp}(n)$ is the empiric mean time to pass level $n$, and $p_c(n)$ the probability to churn at that level. The empiric time to pass $\bar{t}_p^{emp}(n)$ is just the average number of attempts needed by players that passed level $n$ to pass it. The churn probability $p_c(n)$ is the total number of players that abandoned at level $n$ divided by the total number of players that played level $n$.

\begin{figure}[t] 
\includegraphics[width =\linewidth]{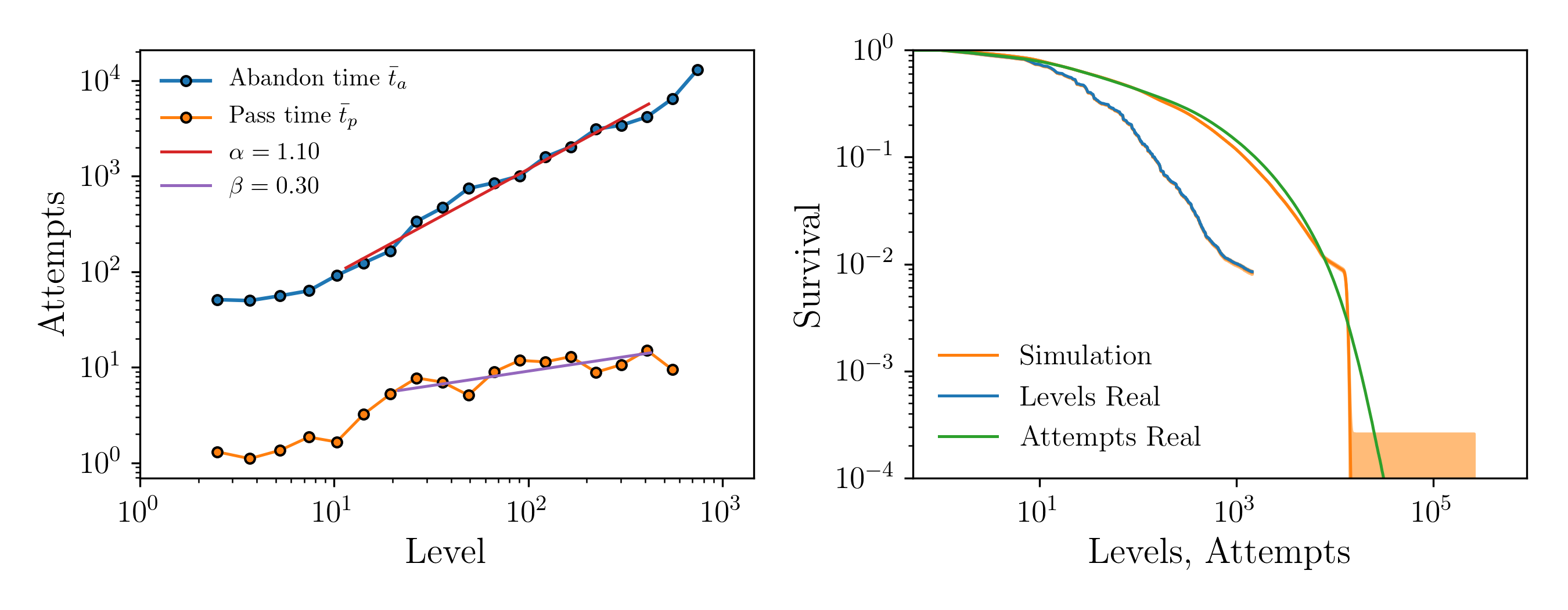}
\caption{Average abandon (in blue) and pass (in orange) times measured at each level of the game Candy Crush Saga from a week cohort of 11,836,502  players with install dates corresponding to the first week of the year 2014 playing on Facebook platform and followed for 2 years. The data has been binned, plotted in double logarithmic scale, and fitted to power laws $\bar{t}_a(n)\sim n^{\alpha}$ and $\bar{t}_p(n)\sim n^{\beta}$, obtaining the values of the exponents $\alpha$ and $\beta$ indicated in the legend. (b) Simulations of the model (orange) reproduce the actual survival of players (blue in levels, green in accumulated attempts) within the statistical uncertainty.}
\label{fig:2}
\end{figure}

{\bf Measuring engagement.} Figure~\ref{fig:2} shows an example of the average abandon and pass times of the different levels of a game, along with the behavior of the survival probability of players in the game as a function of the level and gameplays, respectively. The data corresponds to a week cohort of 11,836,502 players of Candy Crush Saga game playing on the Facebook platform starting on 2014 and followed for 2 years. 

The empirical data reveal a very interesting behavior for the abandon time, $\bar{t}_a(n)$. After an initial number of levels, typically 10-20, where the player is discovering the game (or the activity) and deciding whether he/she likes it or not, the engagement follows a power-law behavior of the form $\bar{t}_a(n)\sim n^{\alpha}$, with an exponent $\alpha$ around 1.1~\footnote{The value of the exponent for most analyzed games is in the range $[1.0,1.5]$. In short datasets there is small plateau for high levels, consequence of the finite time window of observation and the end of content effect (see Fig. S5).}. As a consequence of such fast growth rate, players behave very differently depending on their progression throughout the game, suggesting a "happy-get-happier" mechanism as a final explanation. The average pass time, on the other hand, is an indicator of the relative difficulty of the level as perceived by a player that has reached level $n$ by his/her own means. Therefore, $\bar{t}_p(n)$ is a combination of the intrinsic difficulty of the level and the learning curve of players~\cite{Ritter2001} and, in general, we expect it to show a convex dependency on the progression level $n$. Consider, for instance, the case of learning a musical instrument. It is clear that the Minuet in G (BWV 114) from the Notebook of Anna Magdalena Bach is objectively simpler than the Bach-Brahms Chaconne in D minor BWV 1004 (for the left hand alone). Yet, the effort to learn the former (and so to advance in the progression) is perceived by a first-year piano student as higher than the effort to learn the latter as perceived by, for instance, the great piano player Daniil Trifonov. We thus expect $\bar{t}_p(n)$ to grow with $n$ in a convex way. Our empirical analysis indicates that, indeed, this is the case. As a matter of fact, in the studied datasets, the average pass time after the first 10-20 tutorial levels can be reasonably fitted by a power law $\bar{t}_p(n)\sim n^{\beta}$, with an exponent $\beta$ in the range [0.1,0.5].
\begin{figure}[t] 
\includegraphics[width =\linewidth]{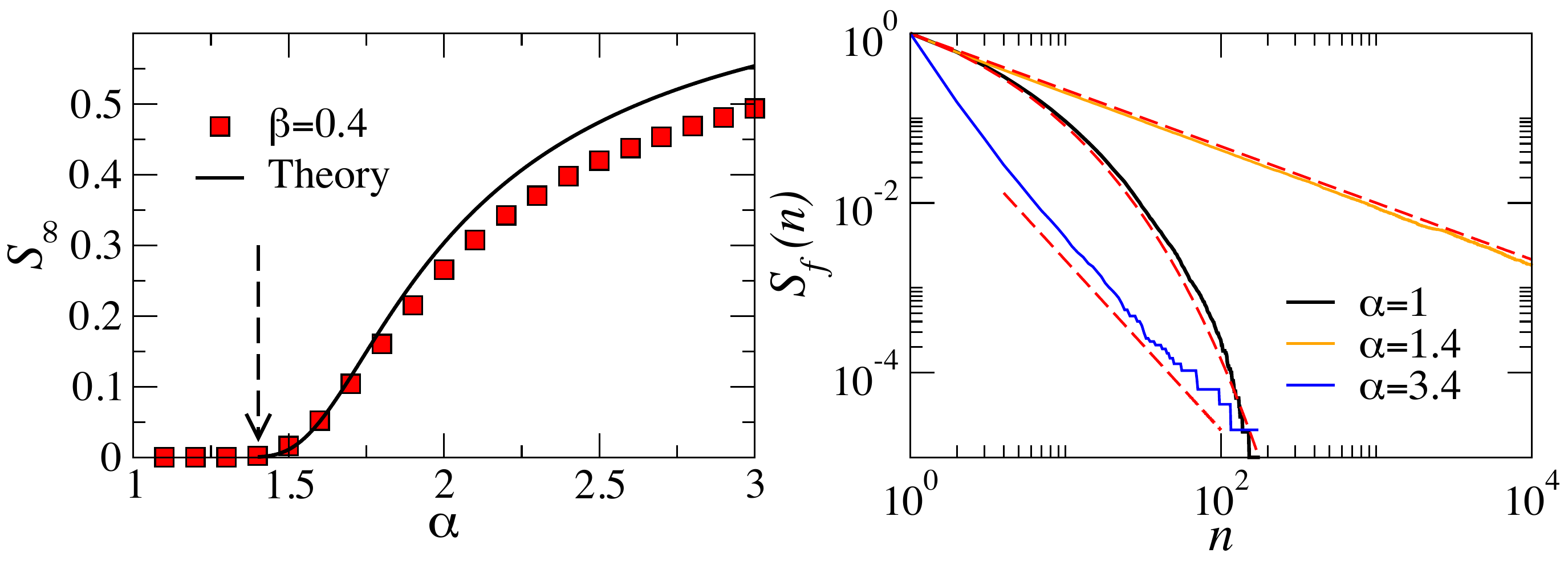}
\caption{Numerical simulations of the phase transition. Left, red squares are results of numerical simulations of the probability of a realization to never end as a function of $\alpha$ for a fixed value of $\beta$. We use the algorithm described in the methods section with $\bar{t}_p(n)=b n^{\beta}, \bar{t}_a(n)=a n^{\alpha}$ with $\beta=0.4$, $a=1.5$, and $b=1$. Solid line is the approximate analytic solution derived in the SI. Notice the smooth approach to the critical point coming from the right, as a consequence of the transition being of infinite order. Right, survival probability of finite realizations below ($\alpha=1$), above ($\alpha=3.4$), and at the critical point ($\alpha=1.4$). Finite realizations are defined as those ending at $n<n_{max}$ with $n_{max}=10^7$. Other values of $n_{max}$ do not change the results significantly. Dashed lines are the analytic predictions.}
\label{fig:3}
\end{figure}

These scaling laws have important consequences for the global dynamics of the game. Indeed, as we show in the SI, there is an (infinite order) phase transition as a function of the parameters $\alpha$ and $\beta$ between a standard phase, where all players eventually quit the game, and an "enthusiastic" phase, where a finite fraction of players never abandon the game. For $\alpha-\beta<1$, the probability of a player quitting the game at level $n$ or higher follows a Weibull distribution of parameter $\beta-\alpha+1$, that is, $S(n)\approx e^{-\mu(n^{\beta-\alpha+1}-1)}$. In this standard phase, the probability of a player to never abandon the game is zero. Instead, when $\alpha-\beta>1$ there is a finite probability that players never abandon the game, provided that the game has infinite content. This probability can be computed as $S_{\infty}\approx e^{\sum_{n=1}^{\infty} \bar{t}_p(n)/\bar{t}_a(n)}$ (see SI for a formal proof). In this "enthusiastic" phase, the survival probability for those players that eventually do abandon the game follows a power law of the form $S_f(n) \sim n^{1+\beta-\alpha}$. This implies that the higher the value of $\alpha-\beta$ the fastest $S_f(n)$ decays, so that either players abandon the game at the beginning of the progression, or they keep playing forever. Interestingly, all the analyzed casual games seem to be below but very close to the critical point $\alpha=1+\beta$ so that the survival probability is well described by a Weibull distribution. Figure~\ref{fig:3}a shows simulation results of this phase transition as compared to the theoretical approximation for $\bar{t}_p(n)=b n^{\beta}$, $\bar{t}_a(n)=a n^{\alpha}$ with $\beta=0.4$, $a=1.5$, and $b=1$. The critical point $\alpha_c=1.4$ and the behavior of $S_{\infty}$ close to the critical point are both very well reproduced by the theoretical approximation. Figure~\ref{fig:3}b shows the survival probability for finite realizations $S_f(n)$ below, at, and above the critical point $\alpha_c$. The agreement with the theoretical predictions is remarkable.

{\bf Mimicking player progression by simulation.} In our model, we make three main assumptions: (i) the independence of the average pass and abandon times; (ii) both times are exponentially distributed; and (iii) all players can be considered as statistically identical. To verify the validity of these assumptions, (i) we performed a detrended fluctuation analysis that verifies that both times are truly independent (see Fig.~S2); (ii) we have also verified that the distribution of abandon and pass times of all levels are exponential to a very good approximation (see Fig.~S3);  (iii) we also show that considering all players as identical reproduces their progression and survival accurately. To contrast the validity of this last assumption and of the model, we simulated the progression and churn of a cohort of identical players using the simple stochastic algorithm described in the Methods section with the abandon and pass times measured for the real dataset as input. Fig.~\ref{fig:2}b compares the real data with the results of the simulations for the survival probability in levels and gameplays (i.e. the fraction of initial players still active after playing a given number of attempts or levels).  The simulations nicely reproduce the real survival (except for the small finite size effects of the tail), showing impressively the validity of the model and of the assumption that all players can be considered as identical. Accordingly, the abandon time is indeed an intrinsic, difficulty-independent measure of the average engagement of players at that level. Hence, a remarkable aspect of the model is that it can measure quantitatively human engagement and how it evolves as players progress in the game.
\begin{figure}[t] 
\includegraphics[width = \linewidth]{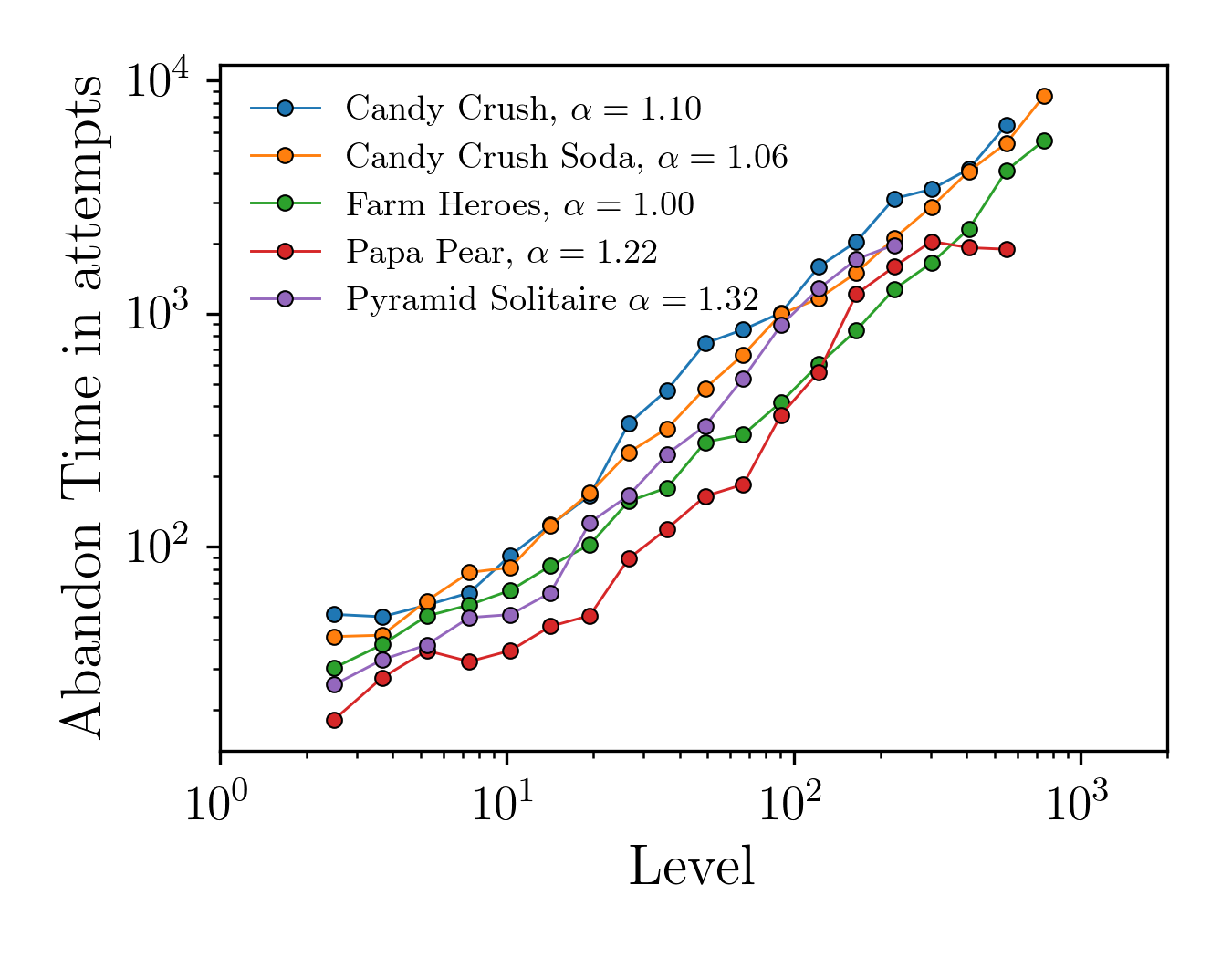}
\caption{Comparison of mean abandon times measured in attempts of different popular Saga games from King: Candy Crush, Candy Crush Soda, Farm Heroes, Papa Pear, and Pyramid Solitaire. In all of them, the abandon time, and thus the engagement, increases as a power-law after the initial 10-20 levels, once players have learnt the dynamics of the game. All data corresponds to a week cohort of installs followed for 2 years. Data from Candy Crush and Pyramid Solitaire are from players on Facebook in the first week of 2014 and the week from 11-10-2013 to 17-10-2013, respectively; the remaining games are from all platforms and the first week of installs in 2017.}
\label{fig:4}
\end{figure}

{\bf Universal Behavior.} The data for the abandon time shown in Fig.~\ref{fig:2} for a specific game (Candy Crush Saga), clearly shows that the engagement increase as a power-law as the player gets more into the game. We repeated the analysis for different Saga games: Farm Heroes, Papa Pear, Candy Crush Soda and Pyramid Solitaire (see Fig.~\ref{fig:4}). These games are very different in terms of genre (e.g. Candy Crush is a match-three swapping tile game; Papa Pear is a physics based bouncing game; Pyramid is a card solitaire), targeted audience, graphics, mechanics and design. Astonishingly, all of them exhibit a common power law behavior of engagement, showing that this evolution of the engagement into a fun activity may be universal. The same happens when we analyzed data corresponding to players from different continents, platforms and periods of time (see Fig.~S4).         

\section*{Discussion} \label{sec:concluding_remarks}

 We have seen that it is possible to quantify and model progression and churn of a playful activity or habit, like a videogame, as a competition between two ingredients: relative difficulty and engagement. Our big data analysis of the system allowed us to find a very precise measure of engagement, which shows a power-law trend indicative of a happy-get-happier mechanism. In this work, we have focused on the particular case of engagement in videogames since, to the best of our knowledge, it is the only system where the amount of available data allows us to elucidate sound statistical laws. However, we believe the process can be generalized to describe engagement in other activities: difficulty is a measure of the training cost and engagement is a measure of the reward or tolerance. Our model shows that a delicate balance between these two ingredients is needed to avoid early churn and that having a very difficult/traumatic experience at the initial stages would lead to massive churn. In addition, there is an interesting phase transition controlled by the ratio of progression between difficulty and engagement that leads to a finite probability that the person never abandons the activity. An interesting example is learning to play a musical instrument and, in general, any rewarding intellectual activity, like doing scientific research or artistic creation. Our model predicts a phase where the probability of individuals to never abandon the activity is non-zero. This may seem as obvious in these cases. Indeed, after many years of intense training, it is very unlikely that a person who had reached an advanced level would stop playing the piano or doing research~\cite{Martin2006}. Certainly, the amount of content in such disciplines is, basically, unlimited and the intellectual reward of keeping doing them is so high that it would be highly improbable that anyone at an advanced level would quit the activity. The importance of our framework relies precisely in its ability to explain when this behavior is possible and under what precise conditions. The model could be helpful to perform a similar analysis in other fields, to quantify tolerance and enjoyment and to design smooth learning procedures to facilitate for instance healthy habits (like sports) or to minimize early school leaving.

\appendix

\section{Empirical estimation of average abandon and pass times of individual levels}

In our model, we assume that pass and abandon times are statistically independent random variables exponentially distributed according to Eq.~\eqref{eq:1}~\footnote{For mathematical tractability we take $t$ as a continuous variable. This assumption does not affect any of the conclusions of this work.}. The corresponding survival probabilities, representing the probability that the time required to pass or abandon at level $n$ is larger than $t$ are:
\begin{equation}
\Psi_n^p(t)=\int_t^{\infty} \psi_n^p(\tau)d\tau=e^{-t/\bar{t}_p(n)}
\end{equation}
and
\begin{equation}
\Psi_n^a(t)=\int_t^{\infty} \psi_n^a(\tau)d\tau=e^{-t/\bar{t}_a(n)}.
\end{equation}
The average abandon, $\bar{t}_a(n)$, and pass, $\bar{t}_p(n)$, times cannot be measured directly from the data. The reason is that abandon and pass times are unconditioned random processes, that is, $\psi_n^p(t)$, for instance, accounts for the distribution of pass times at level $n$ if players were not allowed to quit the game, which is a condition that is not meet in a real dataset. Instead, the empirical observables are: the churn probability at level $n$, $p_c(n)$, defined as the number of players that churned at level $n$ divided by the total number of players that reached that level; and the empirical pass time, $\bar{t}_p^{emp}(n)$, defined as the average time to pass level $n$ for those players that actually passed the level (and, therefore, did not churn). 

In the model, churn probability can be evaluated as the probability that the time to abandon level $n$ --whatever value it takes-- is smaller than the time to pass it. In mathematical terms this is simply expressed as
\begin{equation}
p_c(n)=\int_0^{\infty} \psi_n^a(\tau) \Psi_n^p(\tau) d \tau= \frac{\bar{t}_p(n)}{\bar{t}_p(n)+\bar{t}_a(n)}.
\end{equation}
Similarly, $\bar{t}_p^{emp}(n)$ can be evaluated mathematically in the model as
\begin{equation}
\bar{t}_p^{emp}(n)=\frac{\int_0^{\infty} \tau \psi_n^p(\tau) \Psi_n^a(\tau)d \tau}{\int_0^{\infty} \psi_n^p(\tau) \Psi_n^a(\tau) d \tau}=\frac{\bar{t}_p(n) \bar{t}_a(n)}{\bar{t}_p(n)+\bar{t}_a(n)}.
\end{equation}
By inverting the last two equations, we obtain
\begin{equation}
\bar{t}_p(n)=\frac{\bar{t}_p^{emp}(n)}{1-p_c(n)} \; \mbox{ and } \; \bar{t}_a(n)=\frac{\bar{t}_p^{emp}(n)}{p_c(n)},
\end{equation}
relating the parameters of the model $\bar{t}_p(n)$ and $\bar{t}_a(n)$ with two quantities that can be directly measured in empirical datasets, namely $\bar{t}_p^{emp}(n)$ and $p_c(n)$.

\section{Stochastic Simulations of Player Progression and Churn}

To simulate the model, we only need as input information about $\bar{t}_p(n)$ and $\bar{t}_a(n)$, i.e.  the average time to pass or abandon at level $n$, respectively. In the simulations, for each player starting at $t=1$ at level $n=1$, we perform the following steps~\cite{Erban}:
\begin{enumerate}
\item
Being at level $n$ at time $t$, generate two random numbers $r_1$ and $r_2$, uniformly distributed between $(0,1)$.
\item
Use these random numbers to calculate the time to pass that level as $\tau_p=-\bar{t}_p(n) \ln{r_1}$ and the time to abandon that level as $\tau_a=-\bar{t}_a(n) \ln{r_2}$.
\item
If $\tau_p\le \tau_a$ the player jumps to level $n+1$, time is advanced to $t+\tau_p$, and go to step 1.
\item
If $\tau_p>\tau_a$ the player churns at time $t+\tau_a$ at level $n$.
\end{enumerate}
The whole procedure is then repeated for another player up to a total of $N_1$ players that are used to evaluate the survival curves. The survival curves are calculated as the fraction of the initial number of players that survived up to a given total number of attempts or levels.  The validation of the model was performed using the average abandon and pass time measured from the real dataset and represented in Fig.~\ref{fig:2}a. An excellent agreement was also obtained using the power-law fit as input for the abandon times.

\section{A Continuous Time Random Walk (CTRW) Model of Player Progression and Retention}

Suppose we have a simple linear game where players can access the different levels one by one. The main goal of this model is to evaluate the survival probability of the game $S(t)$, that is, the probability that a given player keeps playing the game after some time $t$, counted from the time the player started playing the game for the first time. In our approach, time is treated as continuous, players are considered as identical and independent, and always progress forward in an increasing manner~\footnote{In real games players that have reached a given maximum level sometimes chose to play lower levels. These events are not very common (e.g. in Papa Pear Saga, less than $2\%$ of gameplays are played at lower levels) and our model ignores them. The process that we model correspond to the stochastic evolution of the maximum level achieved by the player.}. In addition, the assumptions that we make are as follows: 
\begin{enumerate}
\item
When a player reaches a new level $n$, it takes him/her a random time to pass it. This time is controlled by the probability density function (pdf) $\psi^p_n(t)$ that, in general, will depend on the particular level $n$.  
\item
On the other hand, being at level $n$, the player can get bored or frustrated and abandon the game after another random time that follows the pdf $\psi^a_n(t)$, also dependent on the level $n$.
\item
To simplify the model, we assume that these two random times are statistically independent. This means that in order to pass a level, the random time given by the pdf $\psi^p_n$ has to be smaller than the time given by $\psi^a_n$.
\end{enumerate}

The main quantity of interest is the probability that the player is at level $n$ at time $t$, $P_n(t)$. The survival probability of the game can be computed from this distribution as
\begin{equation}
S(t)=\sum_{n=1}^{\infty} P_n(t).
\end{equation}
The probability $P_n(t)$ satisfies the following equation~\cite{Montroll:1965aa}
\begin{equation}
P_n(t)= \int_0^t h_{n}(\tau) \Psi^p_{n}(t-\tau) \Psi^a_{n}(t-\tau) d\tau, \; \; \mbox{for} \; \; n \ge 1
\label{eq:Pndet}
\end{equation}
where $\Psi^p_n(t)$ and $\Psi^a_n(t)$ are the corresponding survival probabilities, that is $\Psi^p_n(t)=\int_t^{\infty} \psi^p_n(\tau) d\tau$ and $\Psi^a_n(t)=\int_t^{\infty} \psi^a_n(\tau) d\tau$, representing the probability that the time required to pass or abandon, respectively at level $n$ is larger than $t$. In turn, $h_n(t)$ is the probability that the player has reached level $n$ between $t$ and $t+dt$, with the initial condition $h_1(t)=\delta(t)$. Eq.~\eqref{eq:Pndet} thus represents the probability that a jump was made to level $n$ at time $\tau\leq t$ and no further transitions to the next level or abandons took place. 
The function $h_n(t)$ satisfies the following self-consistent equation
\begin{equation}
h_n(t)= \int_0^t h_{n-1}(\tau) \psi^p_{n-1}(t-\tau) \Psi^a_{n-1}(t-\tau) d\tau, \; \; \mbox{for} \; \; n \ge 2.
\end{equation}
Notice that the integrals in the last two equations are convolutions, meaning that they can be solved using Laplace transforms. Denoting by $\hat{h}_n(s)$ the Laplace transform of function $h_n(t)$, we can solve it as
\begin{equation}
\hat{h}_n(s)=\prod_{i=1}^{n-1} \mathcal{L}\left\{ \psi^p_i \Psi^a_i \right\}(s) \; \; \mbox{for} \; \; n \ge 2
\end{equation}
where $\mathcal{L}\left\{ \psi^p_i \Psi^a_i \right\}(s)$ denotes the Laplace transform of the product of functions $\psi^p_i(t)$ and $\Psi^a_i(t)$. Using this expression, we can finally write a general formula for the Laplace transform of the survival probability
\begin{equation}
\hat{S}(s)=\mathcal{L}\left\{ \Psi^p_1 \Psi^a_1 \right\}(s)+\sum_{n=2}^{\infty} \mathcal{L}\left\{ \Psi^p_n \Psi^a_n \right\}(s)\prod_{i=1}^{n-1} \mathcal{L}\left\{ \psi^p_i \Psi^a_i \right\}(s).
\label{survival}
\end{equation}
It is quite easy to check the consistency of this expression by considering the case when the player never abandon the game and so $\Psi^a_n(t)=1$ $\forall n$. In such case, the Laplace transform $\hat{S}(s)=1/s$ and, thus, $S(t)=1$.

To make further progress, we need to make some assumptions about the particular form of the probability density functions at each level. We first consider a non-homogeneous Poisson distribution for the abandon time, that is,
\begin{equation}
\Psi^a_n(t)=e^{-k_a(n) t}
\end{equation}
where $k_a(n)$ is the abandon rate, that in general depends on the particular level $n$. Thanks to the properties of the Laplace transform, in this case, the Laplace transform of the product of functions that appears in Eq.~\eqref{survival} is just the Laplace transform of the distributions $\psi^p_n(t)$ but with the argument shifted by a factor $k_a(n)$. Using this property and after some algebra, we can write

\begin{widetext}
\begin{equation}
\hat{S}(s)=\frac{1}{s+k_a(1)}+\sum_{n=1}^{\infty} \frac{k_a(n)-k_a(n+1)}{[s+k_a(n)][s+k_a(n+1)]} \prod_{i=1}^n \hat{\psi^p_i}(s+k_a(i)).
\label{survival2}
\end{equation}
\end{widetext}
Notice that if the abandon rates are independent of the levels, then $k_a(n)=k_a$ and the survival probability is just $S(t)=e^{-k_a t}$, independently of the distributions $\psi^p_n(t)$.  This is easy to understand as in this case the abandon process is a simple homogeneous Poisson process and, thus, independent of the particular levels the player has achieved. Equation~\eqref{survival2} is also interesting because it tells us that in order to have a non trivial result, it is necessary that there is a dependence of the abandon rate on the different levels. The equation is also interesting because by setting $s=0$, we obtain a closed formula for the average survival time $\bar{t}$, which reads
\begin{equation}
\bar{t}=\frac{1}{k_a(1)}+\sum_{n=1}^{\infty} \frac{k_a(n)-k_a(n+1)}{k_a(n) k_a(n+1)} \prod_{i=1}^n \hat{\psi^p_i}(k_a(i))
\label{average_survival}
\end{equation}
which gives us the contribution of each individual level to the overall average survival time of the game. An interesting property made evident by Eq.~\eqref{average_survival} is that flat levels do not contribute to the average survival time. By flat levels we mean sequences of levels with constant abandon rates and so $k_a(n+1) \approx k_a(n)$. That is, a long sequence of similar levels will never increase the average lifespan of players in the game.

\section{Independence of the abandon and pass times}

\begin{figure}[t]
\begin{center}
\includegraphics[width =\linewidth]{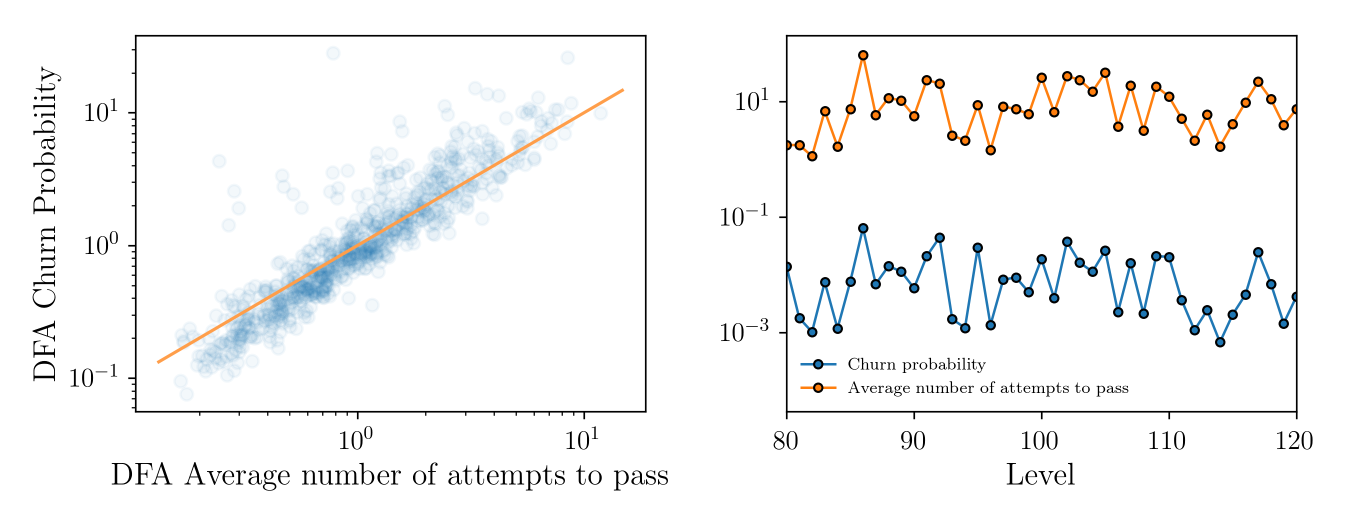}
\caption{Left: Scatter plot showing the detrended correlation between churn probabilities and the average number of gameplays required to pass a particular level for the Candy Crush Saga dataset shown in Fig. 2 of the main text. The orange line indicates the direct proportionality with slope 1. Right: Snapshot of churn probabilities and average number of gameplays to pass a given level as a function of the level. The similarities in both lines provide a clear evidence of the correlation between them. \label{fig:correlations}}
\end{center}
\end{figure}

\begin{figure}[t]
\begin{center}
\includegraphics[width =\linewidth]{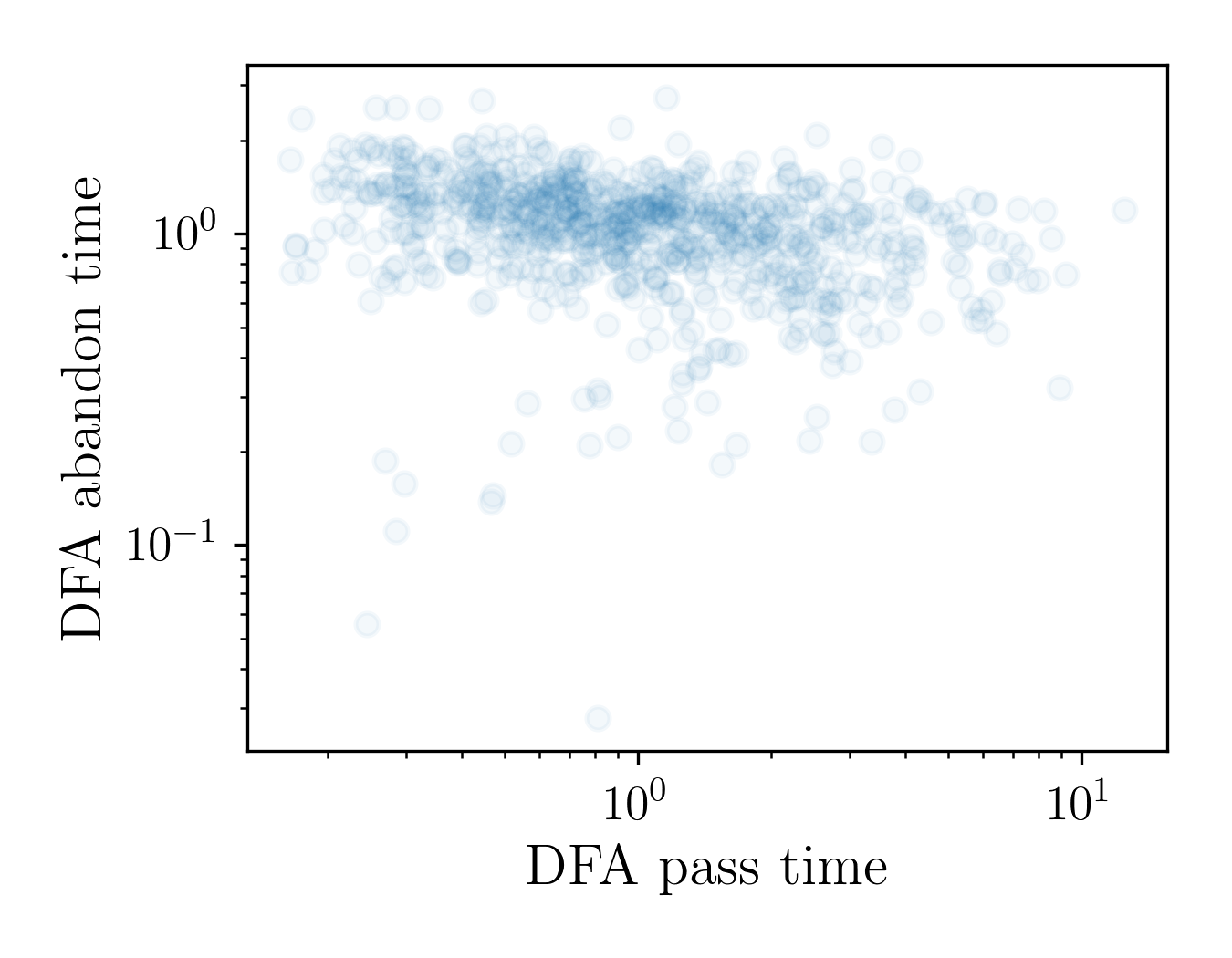}
\caption{A detrended fluctuation analysis of the data in Fig. 2, indeed demonstrates that abandon and pass times are truly uncorrelated.\label{fig:DFA}}
\end{center}
\end{figure}

The previous CTRW model for player progression and churn relies on two main inputs: the probability density distributions of the pass and abandon times, $\psi^p_n(t)$ and $\psi^a_n(t)$. 
For convenience and simplicity, we have assumed that both the pass and abandon times are exponentially distributed, i.e.

\begin{equation}
\psi^p_n(t)= \frac{1}{\bar {t_p}(n) } e^{-t/\bar {t_p}(n)} 
\label{psip}
\end{equation}
\noindent and
\begin{equation}
\psi^a_n(t)=  \frac{1}{\bar {t_a}(n)}e^{-t/\bar {t_a}(n)}
\label{psia}
\end{equation}
\noindent where $\bar {t_p}(n) $ and $\bar {t_a}(n)$ are the average time to pass or abandon at level $n$, respectively. In this case, the average times to pass or abandon at level $n$ are just the inverse of the pass and abandon rates, specifically
\begin{eqnarray}
k_p(n)&=&1/\bar {t_p}(n) \\
k_a(n)&=&1/\bar {t_a}(n).
\end{eqnarray}
\noindent The main parameters of the model, namely the average times to pass, $\bar {t_p}(n)$, or abandon, $\bar {t_a}(n)$, each level $n$, can be directly measured from the datasets in terms of the probability to churn at level $n$, $p_c(n)$, and the empirical time to pass level $n$, $\bar {t_p}^{emp}$, as explained in the methods section. Quite interestingly, these empirical measures provide a strong empirical evidence in favor of our model. We first notice that $\bar {t_p}^{emp}$ and $p_c(n)$ are independent empirical measures. As such, one could have chosen to model the evolution of this process starting directly with these two functions. However, as we show below, both measures are strongly correlated. Interestingly, our CTRW model provides a natural explanation for such correlations. The probability to churn at a given level $p_c(n)$ is typically small, implying that in general we can approximate Eq.~(2) as 
\begin{equation}
p_c(n) \approx \frac{\bar {t_p}(n)}{\bar {t_a}(n)} \; \; \mbox{and} \; \; \bar {t_p}^{emp} \approx \bar {t_p}(n)
\label{pchurn}
\end{equation}
and, therefore, $p_c(n)$ and $\bar {t_p}^{emp}$ should be positively correlated. Figure~\ref{fig:correlations} shows such correlations for the Candy Crush Saga dataset.
 On the other hand, abandon and pass times are assumed to be independent. The validity of this assumption can be tested by analyzing the correlation between both times in the dataset. Fig.~\ref{fig:DFA} shows the results of a detrended fluctuation analysis of the data in Fig.~2, demonstrating that abandon and pass times are indeed truly uncorrelated.

\section{Verification of the exponential behavior of the abandon and pass time distributions}

\begin{figure}[h]
\begin{center}
\includegraphics[width =\linewidth]{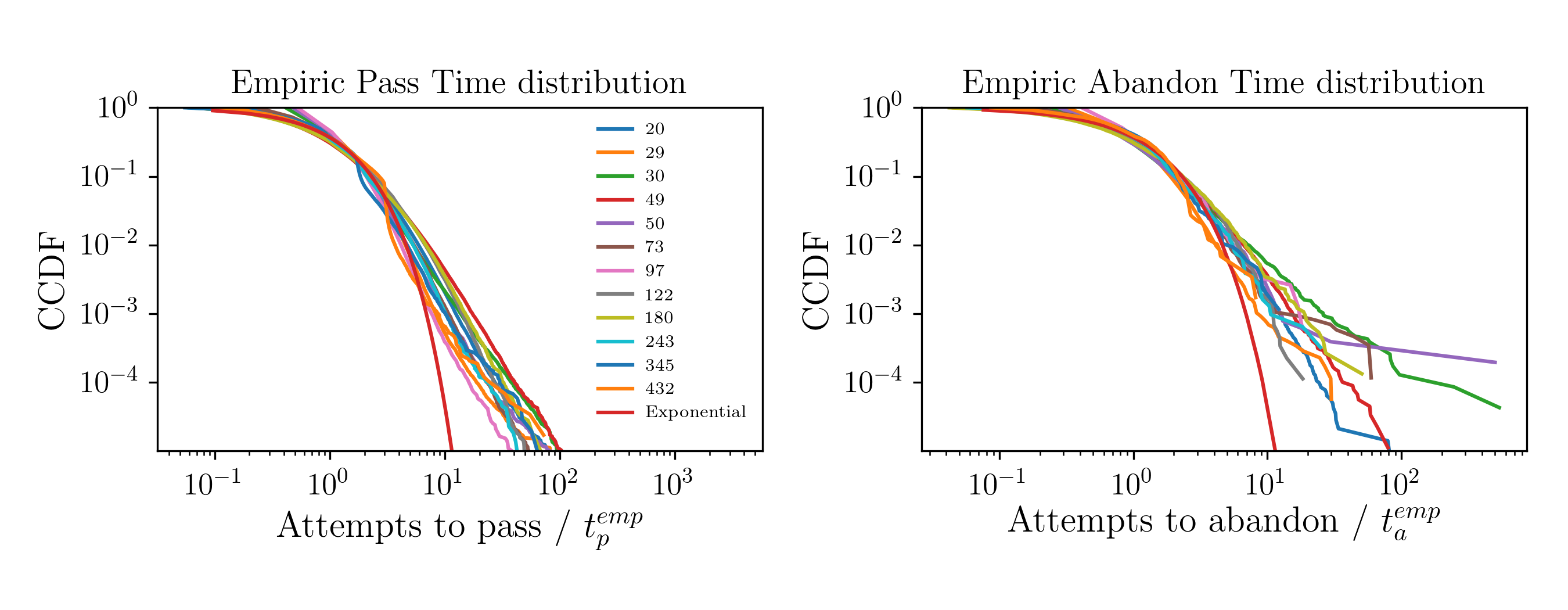}
\caption{Complementary Cumulative Distribution Function (CCDF) of the pass (left) and abandon (right) times, for randomly selected levels of the Blossom Blast Saga game, plotted as a function of the number of attempts divided by the corresponding empiric pass and abandon time. The red line represents in both cases, the expected behavior if the abandon and pass times are exponentially distributed. Data corresponds to a cohort of 4,568,124 players with install dates from 1-1-2016 to 31-1-2016 in all platforms, followed for two years.\label{fig:S3}}
\end{center}
\end{figure}

From the dataset, one cannot measure directly the probability distribution function of abandon, $\psi^a_n(t)$, and pass times, $\psi^p_n(t)$ of a specific level. This is due to the fact that abandon and pass times are unconditioned random processes. That is,  $\psi_n^p\left(t\right)$ accounts for the distribution of pass times at level $n$ if players were not allowed to quit the game, which is a condition that is not meet in a real dataset. Similarly, $\psi_n^a\left(t\right)$ is the distribution of abandon times at level $n$ if players were not allowed to quit the game.
Instead, the distributions that we can observe directly are the {\it empiric } distribution of pass and abandon times. These distributions can be simply obtained as the normalized histogram of the attempts required to pass or abandon a specific level, and are mathematically given by
\begin{equation}
 \bar{\psi}_n^p\left(t\right)=\frac{\psi_n^p\left(t\right)\psi_n^a\left(t\right)}{\int_0^\infty{\psi_n^p\left(t\right)\psi_n^a\left(t\right)}}
\end{equation} 

\begin{equation}
\bar{\psi}_n^a\left(t\right)=\frac{\psi_n^a\left(t\right)\psi_n^p\left(t\right)}{\int_0^\infty{\psi_n^a\left(t\right)\psi_n^p\left(t\right)}}
\end{equation} 
The right hand side of the previous equations represents the distribution of pass times conditioned to the fact that the player has not yet churned, and the distribution of abandon times conditioned to the fact that the player has not yet passed level $n$ at time $t$, respectively. 
In the case that the unconditional probabilities are exponentially distributed as in Eqs.~\ref{psip} and \ref{psia}, it is easy to show that the Complementary Cumulative Distribution Function (CCDF) of the empiric distributions of pass and abandon times is just given by
\begin{equation}
 \bar{\Psi}_n^p\left(t\right)=\bar{\Psi}_n^a\left(t\right)= e^{-t/\bar{t}^{emp}(n)}
\end{equation}
\noindent where $\bar{t}^{emp}(n)=\bar{t}_{p}(n)+\bar{t}_{a}(n)$.
Fig. \ref{fig:S3} represents the CCDF for different levels of Blossom Blast Saga, plotted as a function of the number of attempts divided by the corresponding empiric mean time.

The CCDF distribution for all levels nicely collapses in a single master curve that is very well approximated by the predicted exponential behavior, demonstrating that, to a good approximation, the distribution of abandon and pass times are indeed exponential.

\begin{figure*}[t]
\begin{center}
\includegraphics[width =\linewidth]{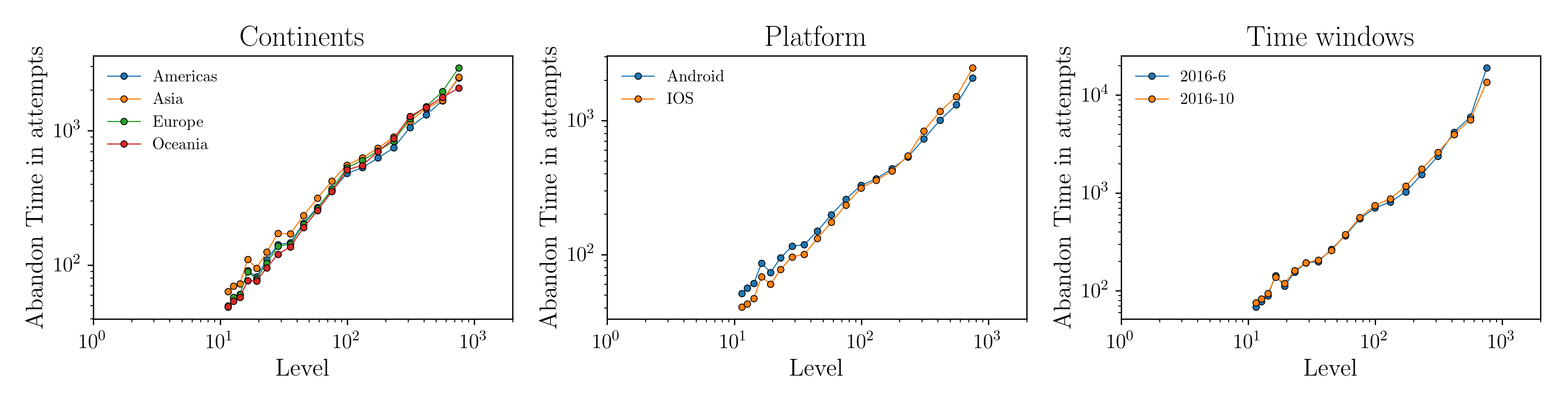}
\caption{Average abandon times measured at each level of the game Candy Crush Soda Saga from all players  in the period 1-06-2016 to 31-07-2016. The data has been binned and plotted in double logarithmic scale. Left: Abandon time measured for players segmented according to their continent. Middle: Abandon time measured for players using Android or OS as platform. Right: Abandon time measured for different time periods, corresponding to June and October 2016.
In all cases, a clear power-law behavior is observed.\label{fig:S4}}
\end{center}
\end{figure*}
\section{``Universality'' of the power-law dependence of the abandon times}

We have measured from different datasets the abandon and pass times of each individual level for players of different continents, playing using a different platform, and that have installed the game and are playing at different periods of time intervals. In all cases, for each game we obtained almost identical average pass times (not shown) and a consistent power-law behavior of the average abandon times with very similar exponents (see Fig.~\ref{fig:S4}). This is a clear indication of the ``universal'' power-law behavior of the engagement in this fun activity.

\section{Finite size effects}
\begin{figure}[t]
\begin{center}
\includegraphics[width =\linewidth]{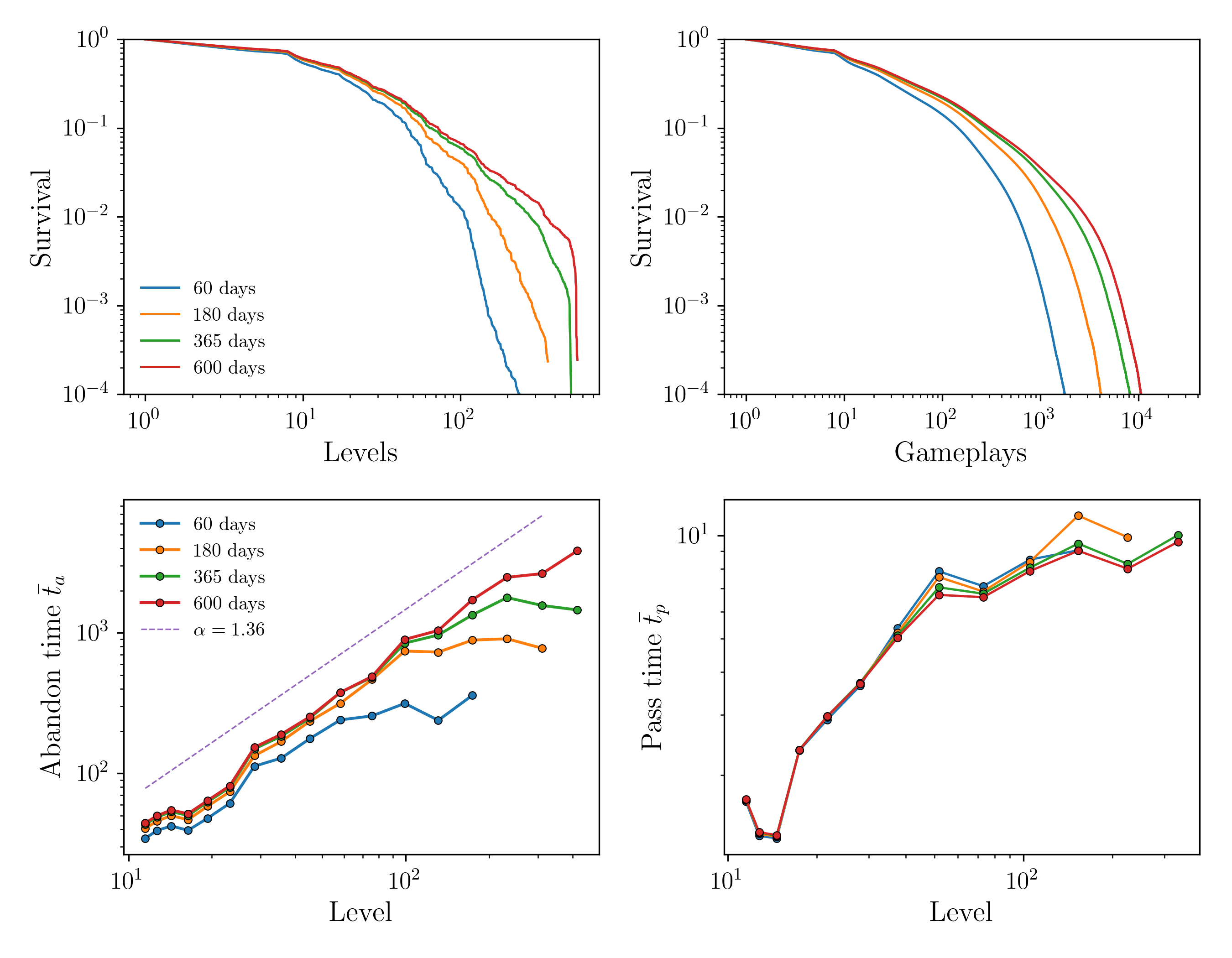}
\caption{ Illustration of the finite time window effects. The plots represent the change in the fraction of players still active in the game after a total number of gameplays (a) or levels (b), the mean abandon (c) and pass times (d), for Papa Pear Saga dataset on the Facebook interface of a week cohort of users with installation date from 11/10/2013 to 18/10/2013 measured using different intervals of real time activity, spanning from 2 to 24 months. For short time windows, most players have not had the time to play high levels and this leads to a clear cut-off in the survival curves and a plateau in the observed abandon times.
\label{fig:FSS}}
\end{center}
\end{figure}

The estimation of average abandon and pass times are affected by the length of the dataset, that is, the time span during which we follow our cohort of players. This is so because empirically we consider that a player has abandoned the game at his/her last observed gameplay. However, if we increase the observation time window, some players may be still active in the game even thought they were considered as non-active with the smallest time window. This affects the estimation of $p_c(n)$ and, thus, of $\bar {t_p}(n)$ and $\bar {t_a}(n)$. These effects are more evident in Fig.~\ref{fig:FSS} comparing how the number of alive players after a given number of gameplays or levels, and the abandon and pass times change with the period of time used in their evaluation. Whereas the pass times seem to be quite stable, the tail of the abandon times is strongly affected by data-censorship due to the finite time window of analysis. However, as the time window increases, we observe a clear collapse towards a clean power law behavior.

\section{Phase transition}

The model undergoes a phase transition between a phase where all players eventually quit the game and a phase where a finite fraction of players never abandon the game. The probability of a player to be still playing at level $n$ is simply the probability of not having churned in any level below $n$, that is,
\begin{equation}
S(n)=\prod_{i=1}^n (1-p_c(i)).
\end{equation}
For all levels, $p_c(n)$ is always a small number and, therefore, we can approximate this expression as
\begin{equation}
S(n) \approx e^{-\sum_{i=1}^n p_c(i)} \approx e^{-\int_{i=1}^n p_c(i)di}
\end{equation}
where in the last approximation we have taken the continuum approximation. Using Eq.~\eqref{pchurn} and assuming that $\bar{t}_a(n)=a n^\alpha$ and $\bar{t}_p(n)=b n^\beta$, $S(n)$ can be expressed as
\begin{widetext}
\begin{equation}
S(n)\approx e^{-\frac{b}{a} \int_1^n i^{\beta-\alpha}di}=\left\{
\begin{array}{lr}
e^{-\frac{b}{a(\beta-\alpha+1)}(n^{\beta-\alpha+1}-1)} & \alpha \ne 1+\beta\\[0.5cm]
\frac{1}{n^{b/a}} & \alpha=1+\beta
\end{array}
\right. .
\end{equation}
\end{widetext}
When $\alpha<1+\beta$, the limit $\lim_{n\rightarrow \infty} S(n)=0$, which implies that all players eventually abandon the game. However, when $\alpha>\alpha_c=1+\beta$, the survival probability $S(n)$ converges to a constant value. Therefore, in this case, there is finite probability that a player never abandon the game $S_{\infty}$ given by
\begin{equation}
S_{\infty}=e^{-\frac{b}{a(\alpha-\alpha_c)}}.
\end{equation}
Notice that $S_{\infty}$ and all its derivatives of any order vanishes at $\alpha=\alpha_c$ (evaluated from the right) so that the phase transition is of infinite order. When $\alpha>\alpha_c=1+\beta$, the survival probability of players with finite lifespan can be evaluated as
\begin{equation}
S_{fin}(n)=\frac{S(n)-S_{\infty}}{1-S_{\infty}},
\end{equation}
that, for $n\gg 1$ behaves as $S_{fin}(n) \sim n^{\alpha_c-\alpha}$.

The interpretation is then as follows: for $\alpha \ll \alpha_c$ players' lifespans are short. When $\alpha \approx \alpha_c$ from below, the average lifespan grows and diverges right at the critical point, even though all players eventually abandon the game. Above the critical point $\alpha>\alpha_c$, there is a fraction of players that never abandon the game, and those that do abandon the game follow a power law distribution with exponent $\alpha-\alpha_c+1$. When $\alpha \gg \alpha_c$, players either stay in the game forever or have a very short lifespan, abandoning the game at very low levels.

%\bibliographystyle{plain}
%\bibliography{paper}

%
\end{document}